\documentclass[singlespacing]{elsart}
\usepackage{epsfig}
\begin{document}
\begin{frontmatter}

\title{Low Energy Singlets in the Excitation
Spectrum of the Spin Tetrahedra System $\bf Cu_2Te_2O_5Br_2$}

\author[label0,label1]{ P. Lemmens}\ead{peter.lemmens@physik.rwth-aachen.de}\ead[url]{www.peter-lemmens.de}
\author[label1]{ K.Y. Choi}\author[label1]{ A. Ionescu}
\author[label1]{ J. Pommer}\author[label1]{ G. G{\"u}ntherodt}\author[label6]{ R. Valent\'\i}
\author[label6]{ C. Gros}\author[label2]{ W. Brenig} \author[label3]{ M. Johnsson}
\author[label4]{ P. Millet}\author[label5]{F. Mila}

\address[label0]{IMNF, TU Braunschweig, D-38106 Braunschweig, Germany}
\address[label1]{2. Phys. Inst., RWTH Aachen, D-56056 Aachen, Germany}
\address[label6]{Inst. f{\"u}r Theor. Physik, Univ. des Saarlands, D-66041 Saarbr{\"u}cken,
Germany}
\address[label2]{Inst. f{\"u}r Theor. Physik, TU Braunschweig, D-38106 Braunschweig,
Germany}
\address[label3]{Dept. of Inorganic Chem., Stockholm Univ., S-10691 Stockholm,
Sweden}
\address[label4]{CEMES/CNRS, F-31062 Toulouse, France}
\address[label5]{Inst. de Phys. Th\'eorique, Univ. Lausanne, CH-1015 Lausanne,
Switzerland}

\begin{abstract}
Low energy Raman scattering of the s=1/2 spin tetrahedra system
$\rm Cu_2Te_2O_5Br_2$ is dominated by an excitation at $\rm
18~cm^{-1}$ corresponding to an energy $\rm E_S=0.6\Delta$, with
$\rm \Delta$ the spin gap of the compound. For elevated
temperatures this mode shows a soft mode-like decrease in energy
pointing to an instability of the system. The isostructural
reference system $\rm Cu_2Te_2O_5Cl_2$ with a presumably larger
inter-tetrahedra coupling does not show such a low energy mode.
Instead its excitation spectrum and thermodynamic properties are
compatible with long range N\'{e}el-ordering. We discuss the observed
effects in the context of quantum fluctuations and competing
ground states.
\end{abstract}

\end{frontmatter}

\section{Introduction}
\label{intro} Raman light scattering is a powerful tool to
investigate the low energy excitation spectrum of quantum spin
systems realized, e.g. in transition metal oxides. This is based
on its high sensitivity to all relevant degrees of freedom,
especially to collective magnetic excitations as singlet modes or
bound states, that are not observable in neutron scattering
experiments. Singlet modes play an essential role in the
description of many particle effects in the gapfull spin liquid
phase of quantum spin systems \cite{mila00}. Furthermore,
scenarios have been developed for systems that are at a boundary
to another competing phase, e.g. related to long range magnetic
ordering \cite{kotov01}. Interesting in this respect are
frustrated magnets like the Shastry-Sutherland lattice
\cite{lemmens00} or the $\rm Kagom\acute{e}$-lattice
\cite{ramirez00}. In this report we demonstrate, that a recently
found quantum spin system based on weakly coupled $\rm
Cu^{2+}$-spin tetrahedra, $\rm Cu_2Te_2O_5Br_2$ \cite{johnsson00},
shows low energy modes related to the spin system and an unusual
instability that is easily tunable by the composition and the
volume of the unit cell, i.e. a substitution of Br- by the smaller
Cl-anions.

\section{Results and Discussion}
The crystal structure of $\rm Cu_2Te_2O_5Br_2$ is shown in Fig.~1.
It consists of $\rm Cu^{2+}$-spin tetrahedra that are formed by an
O- and Br-superexchange network of ions that is additionally
supported by the {\it lone pair}-cation Te. Magnetic
susceptibility measurements show a maximum at $\rm
T_{\chi_{max}}$~=~23~K for this compound and at 30~K for the
isostructural $\rm Cu_2Te_2O_5Cl_2$. A strong reduction is evident
at low temperatures, consistent with a spin gap system. A fit to
these susceptibility data has been obtained with an
intra-tetrahedra coupling of $\rm J\approx$ 40~K and a spin gap
$\Delta$ of similar magnitude \cite{johnsson00}.

The excitation spectrum of such an individual tetrahedron is
simple, however, the inherent degeneracy in the singlet sector may
lead to interesting low temperature properties. The 16 states
divide into a quintuplet, two triplets (with one state doubly
degenerate) and two singlet states. The later singlets should form
the ground state of the magnetic system. A seemingly negligible
distortion of the tetrahedra should prefer one of them and shift
the other to a small but higher energy. This low energy excited
singlet state above the singlet ground state should only be
observable in magnetic Raman scattering due to $\rm \Delta S=0$
Heisenberg exchange scattering.

In Fig.~2 Raman scattering spectra in (cc) polarization of the
bromide and the chloride are compared at low temperatures. In $\rm
Cu_2Te_2O_5Br_2$ a pronounced mode is evident and attributed to a
singlet state at $\rm E_S=18~cm^{-1} \approx 0.6\Delta$. In
addition, a pyramidal-shaped scattering continuum is observed
centered at 61~$\rm cm ^{-1}$~=~88~K with a total linewidth of
40~$\rm cm ^{-1}$. This continuum is attributed to a
two-magnon-like scattering process \cite{brenig01} and its
linewidth points to an appreciable inter-tetrahedra coupling. The
polarization selection rule on the other hand supports a
predominant exchange path along the crystallographic c-axis of the
compound. This would justify models of 1D coupled tetrahedra
\cite{brenig01}. The center energy corresponds very well to $\rm
2\Delta~\approx~80~K$ determined from the magnetic susceptibility
\cite{johnsson00}. In the chloride the low energy scattering is
less pronounced and even more spread out. Also the phonon modes
differ in frequency due to the different volume of the unit cell
of the two compounds.

A study of the temperature dependence of the spectra in $\rm
Cu_2Te_2O_5Br_2$ shows a strong soft mode-like decrease of the
frequency of the mode at $\rm E_S=18~cm^{-1}$ pointing to some
instability. Thermodynamic experiments indeed support this
evidence since a broad bump at $\rm T_o$=11.4~K is observed in the
specific heat together with a kink in the magnetic susceptibility
\cite{lemmens01}. The transition temperature shows a well
pronounced increase with a magnetic field. In contrast, for $\rm
Cu_2Te_2O_5Cl_2$ the specific heat shows a lambda-like anomaly at
$\rm T_N$=18.2~K and essentially no effect of a magnetic field.
Here, all observations are consistent with long-range $\rm
N\acute{e}el$-Ordering.

\section{Conclusion}
The peculiar experimental observations in these two Cu-Oxo-Halides
give evidence that, although structurally related, their low
energy excitation spectrum and ground state properties are very
distinct. Although details of the electronic band structure,
hopping elements and especially the important inter-tetrahedra
exchange coupling constants of these systems are unknown, some
conclusions can be made. In first approximation and neglecting the
electronic difference between Cl- and Br-anions the main
difference is the 7\% larger unit cell volume of $\rm
Cu_2Te_2O_5Br_2$, that should be related to a smaller
inter-tetrahedra coupling. This coupling is evidenced in a more
concentrated spectral weight of the two-magnon-like continuum in
Raman scattering and the smaller transition temperature. In the
proposed scenario the transition itself lifts the degeneracy of
the singlet states. This however does not necessarily mean that a
large structural distortion is involved. A careful investigation
of the optical phonon excitations does not give any evidence for
superlattice peaks. In this sense the phenomenology is much
different from a spin-Peierls transition, as observed in $\rm
CuGeO_3$, or the charge ordering related instability in $\rm
NaV_2O_5$, although in these compounds low energy singlets are
observed as well \cite{muthu,lemmens-navo}. Further studies are
underway to understand this material.

{\bf Acknowledgement:} We are grateful for helpful discussions with H.
Kageyama, Yu. Ueda, and K. Ueda. This work was supported in part by the
DFG through SPP1073 and the Swiss Science Foundation under grant nr.
21-63749.

\newpage
\begin{figure}
\begin{center}
\centerline{\epsfig{figure=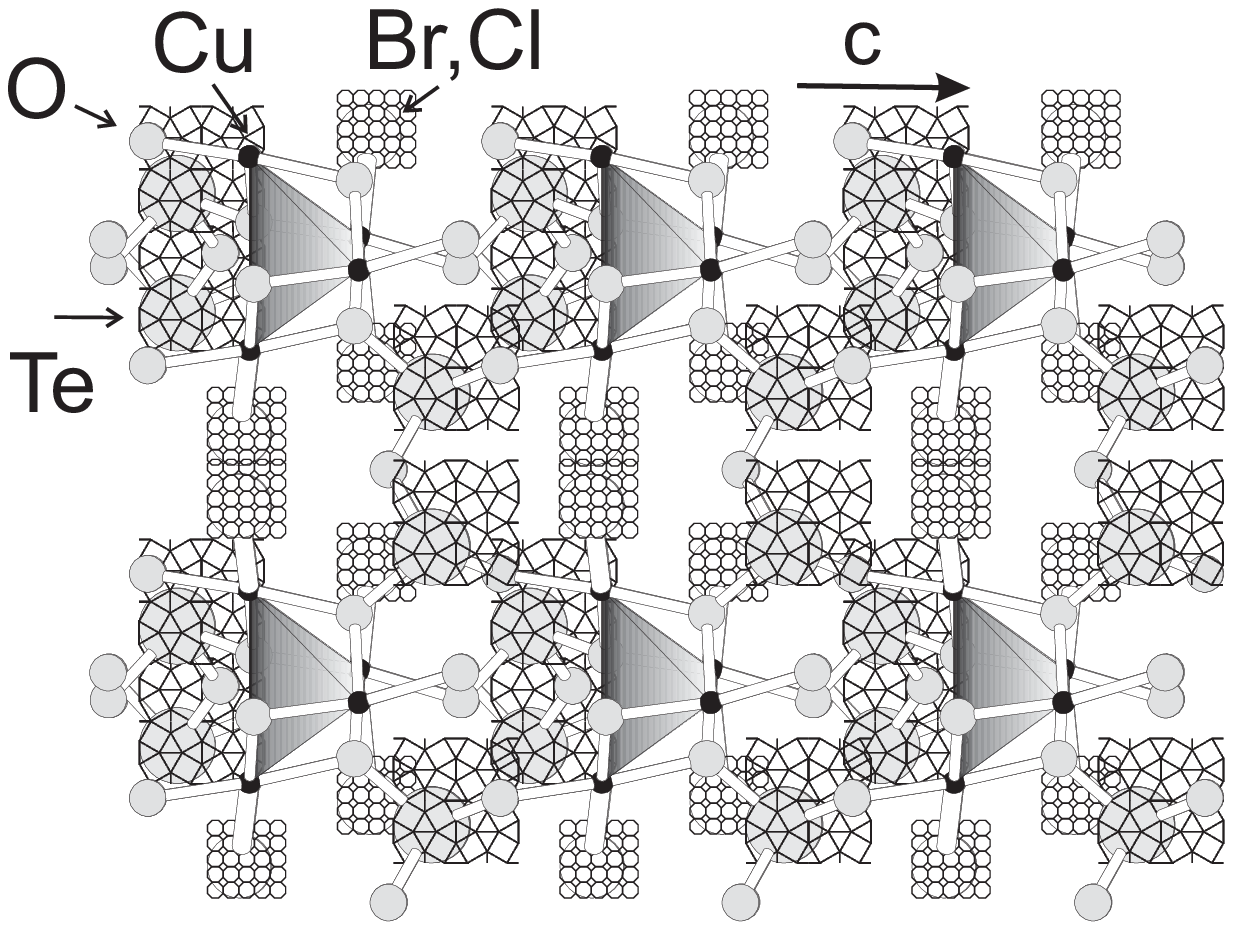,width=10cm}}
\vspace{1cm}\hspace{5cm} \caption{Crystal structure of $\rm
Cu_2Te_2O_5X_2$, X=Br or Cl, with hatched Cu-tetrahedra. }
\end{center}
\end{figure}

\begin{figure}
\begin{center}
\centerline{\epsfig{figure=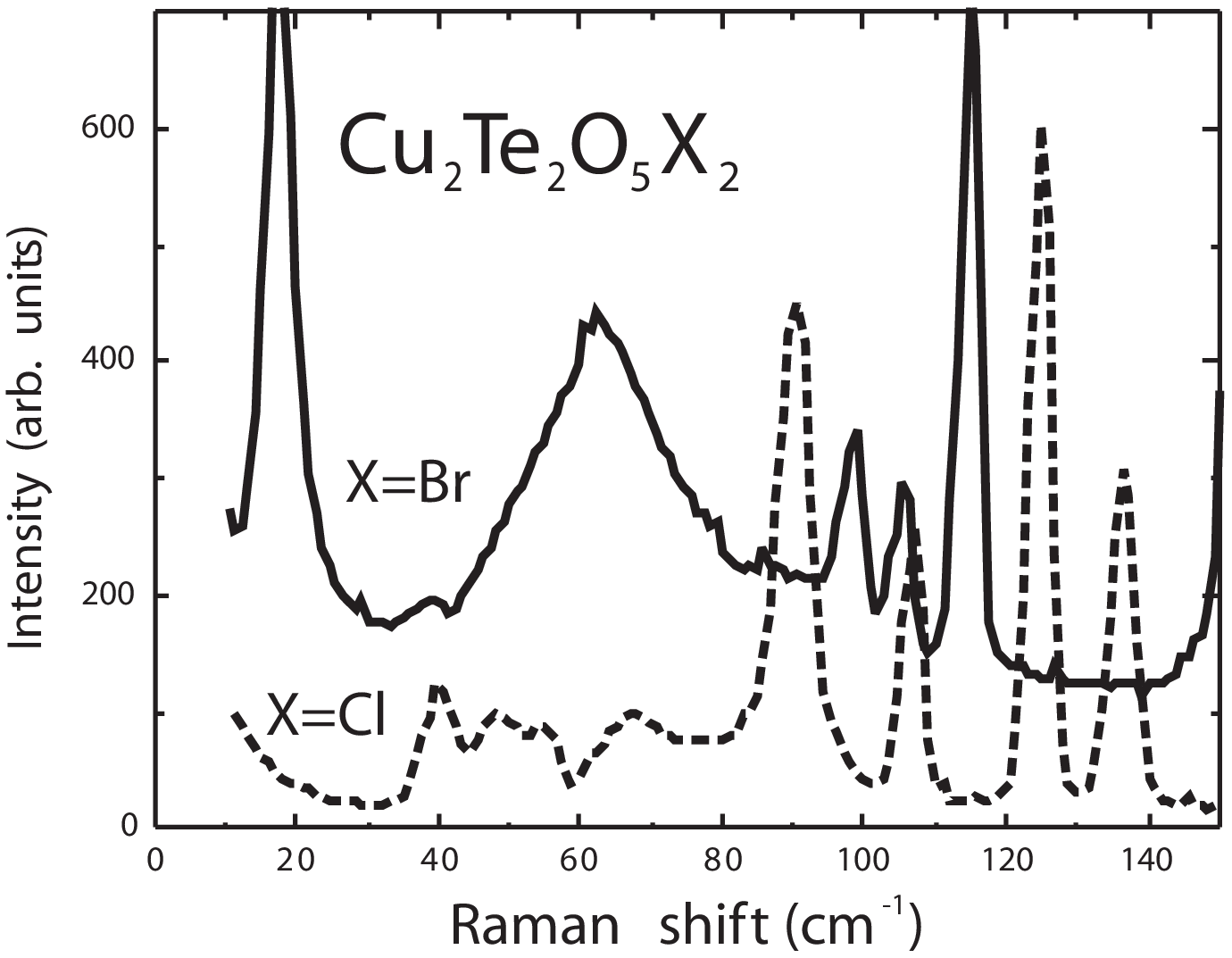,width=10cm}}
\vspace{1cm}\hspace{5cm} \caption{Raman spectra of $\rm
Cu_2Te_2O_5Br_2$ and the isostructural $\rm Cu_2Te_2O_5Cl_2$ at
low temperatures (T=3K) in c-axis polarization (cc). The spectrum
of $\rm Cu_2Te_2O_5Br_2$ has been shifted upwards by 100 counts.}
\end{center}
\end{figure}

\end{document}